\documentstyle[12pt,epsfig]{article}
\begin{document}

\vspace{5mm} 
\begin{center} 
{\Large \bf QCD prerequisites for extra dimension searches }\\
\vspace{5mm} 
{\bf 
Prakash Mathews$^{a,}$\footnote{Invited talk presented at the XVII DAE-BRNS
Symposium on High Energy Physics, IIT Kharagpur, 11-15 December 2006.}, 
V. Ravindran$^b$ 
}\\ 
\end{center}
\vspace{10pt} 
\begin{flushleft}
{\it 
a) Saha Institute of Nuclear Physics, 1/AF Bidhan Nagar,
Kolkata 700 064, India\\
\vspace{3pt} 
\hspace{1cm}{\sf E-mail: ~prakash.mathews@saha.ac.in}
\vspace{10pt} 
 
b) Harish-Chandra Research Institute, 
 Chhatnag Road, Jhunsi, Allahabad, India\\
\vspace{3pt} 
\hspace{1cm}{\sf E-mail: ~ravindra@mri.ernet.in}
} 
\end{flushleft}
 
\vspace{10pt} 
\noindent
{\bf Abstract} 
\vskip 0.3 cm

For the dilepton production at hadron collider in TeV scale gravity models,
inclusion of QCD corrections to NLO stabilises the cross section with 
respect to scale variations.  The K-factors for the various distributions
for the ADD and RS model at both LHC and Tevatron are presented.

\vskip12pt 

\section{Introduction}

Extra dimension scenarios are now essential part of the studies of
physics beyond the Standard Model (SM).  They provide an alternate
view of the hierarchy between the electroweak and the Planck
scale.  Some of these extra dimension models invoke the brane world
scenarios to hide the extra spacial dimensions from current observation.
Two such models that are phenomenologically widely studied are the
Arkani-Hamed, Dimopoulos and Dvali (ADD) \cite{add} and the
Randall-Sundrum (RS) \cite{rs} models.

In the ADD case the compactified extra dimensions could be large and the
large volume of the compactified extra spacial dimension would account for
the dilution of gravity in 4-dimensions and hence the hierarchy.  In this
model, new physics can appear at a mass scale of the order of a TeV.  A
viable mechanism to hide the extra spacial dimension, is to introduce a
3-brane with negligible tension and localise the SM particles on it.  Only
the graviton is allowed to propagate the full $4+d$ dimensional space
time.  As a consequence of these assumptions, it follows from Gauss Law
that the effective Planck scale $M_P$ in 4-dimension is related to the
$4+d$ dimensional fundamental scale $M_S$ through the volume of the
compactified extra dimensions \cite{add}.  The extra dimensions
are compactified on a torus of common circumference $R$.  
The space time is
factorisable and the 4-dimensional spectrum consists of the SM confined to
4-dimensions and a tower of Kaluza-Klien (KK) modes of the graviton propagating
the full $4+d$ dimensional space time.
The interaction of the KK modes $h_{\mu\nu}^{(\vec n)}$ with the SM fields
localised on the 3-brane is given by
\begin{eqnarray}
{\cal L}_{int} \sim - \frac{1}{M_P} \sum_{\vec n=0}^\infty T^{\mu\nu} (x)
                      h_{\mu\nu}^{(\vec n)} (x) ~,
\end{eqnarray}
where $T^{\mu\nu}$ is the energy-momentum tensor of the SM fields on the
3-brane.  The zero mode corresponds to the usual 4-dimensional massless
graviton.  The KK modes are all $M_P$ suppressed but the high multiplicity
could lead to observable effects at present and future colliders.

In the RS model there is only one extra spacial dimension and the extra
dimension is compactified to a circle of circumference $2 L$ and further
orbifolded by identifying points related by $y \to -y$.  Two branes are
placed at orbifold fixed points, $y=0$ with positive tension called the
Planck brane and a second brane at $y=L$ with negative tension called the
TeV brane.  For a special choice of parameters, it turns out that the
5-dimensional Einstein equations have a warped solution for $0<y<L$ with
metric $g_{\mu\nu} (x^\rho,y)=\exp(-2 k y) ~\eta_{\mu\nu}$, $g_{\mu y}=0$
and $g_{y y}=1$.  This space is not factorisable and has a constant negative
curvature--- $AdS_5$ space-time.  $k$ is the curvature of the $AdS_5$
space-time and $\eta_{\mu\nu}$ is the usual 4-dimensional flat Minkowski
metric.  In this model the mass scales vary with $y$ according to the
exponential warp factor.  If gravity originates on the brane at $y=0$,
TeV scales can be generated on the brane at $y=L$ for $k L \sim 10$.
The apparent hierarchy is generated by the exponential warp
factor and no additional large hierarchies appear.  The size of the
extra dimension is of the order of $M_P^{-1}$.  Further it has been
showed that \cite{gw} the value of $k L$ can be stabilised without
fine tuning by minimising the potential for the modulus field
which describes the relative motion of the two branes.  In the RS
model graviton and the modulus field can propagate the full
5-dimensional space time while the SM is confined to the TeV brane.
The 4-dimensional spectrum contains the KK modes, the zero mode is
$M_P$ suppressed while the excited modes are massive and are only TeV
suppressed.  The mass gap of the KK modes is determined by the difference
of the successive zeros of the Bessel function $J_1 (x)$ and the scale
$m_0=k ~e^{-\pi k L}$.
As in the ADD case the phenomenology of the RS model concerns
the effect of massive KK modes of the graviton, though the spectrum
of the KK mode is quite different.
In the RS model the massive KK modes $h_{\mu\nu}^{(n)} (x)$ interacts
with the SM fields
\begin{eqnarray}
{\cal L}_{int} \sim - \frac{1}{M_P} T^{\mu\nu} (x) h_{\mu\nu}^{(0)} (x)
                 - \frac{e^{\pi k L}}{M_P} \sum_{n=1}^\infty T^{\mu\nu} (x)
                   h_{\mu\nu}^{(n)} (x) ~.
\end{eqnarray}
The masses of $h_{\mu\nu}^{(n)} (x)$ are given by $M_n=x_n
~k ~e^{-\pi k L}$, where $x_n$ are the zeros of the Bessel function $J_1
(x)$.  In the RS model there are two parameters which are $c_0=k/M_P$, the
effective coupling
and $M_1$ the mass of the first KK mode.  Expect for
an overall warp factor the Feynman rules of RS are the same as those of the
ADD model.

These models are being tested at the Tevatron \cite{d01,d02} and will be 
tested at the LHC in the near future.  
One of the most important channels that could probe the signatures of these
extra-dimensional models is large invariant mass di-lepton pair production,
(also called Drell-Yan(DY) production)
in hadronic colliders.  In this process, the gravitons which are KK modes
of the theory can appear in the intermediate state to produce di-lepton
pairs because gravity couples to anything and everything.  In ADD model, even though
the coupling of gravity to standard model particles is small, the
collective effect coming from all the KK modes can result in observable effect
in some kinematic domains that can be probed.  In RS, the warp factor
compensates the small coupling making the effect feasible.  Since the
process under study is initiated by quarks and gluons, there is a large
theoretical uncertainty coming from factorisation and renormalisation scales
and also from missing higher order corrections resulting from strong interaction
dynamics namely Quantum Chromodynamics(QCD).  It is important for such searches 
to have some quantitative estimate of the effects of higher order QCD 
corrections
and here we review the recent results on the NLO QCD corrections for various 
distributions of the di-lepton pair production both at the Tevatron and LHC.


\section{ Di-lepton pair production at hadron collider} 

Consider the collision of hadrons $P_1,P_2$ to leptonic final states
$P_1(p_1)+P_2(p_2) \rightarrow \ell^+(l_1)+\ell^-(l_2)+X(P_X)$,
where $X$ denotes the final inclusive hadronic state and carries the 
momentum $P_X$.  In the QCD improved parton model, the hadronic cross 
section can be expressed in terms of partonic cross sections
convoluted with appropriate parton distribution functions
\begin{eqnarray}
2 S~{d \sigma^{P_1 P_2} \over d Q^2}\left(\tau,Q^2\right)
&=&\sum_{ab={q,\overline q,g}} \int_0^1 dx_1
\int_0^1 dx_2~ f_a^{P_1}(x_1) ~
f_b^{P_2}(x_2)
\nonumber \\[2ex] 
&&
\times \int_0^1 dz \,\, 2 \hat s ~
{d \hat \sigma^{ab} \over d Q^2}\left(z,Q^2\right)
\delta(\tau-z x_1 x_2)\,.
\label{eq8}
\end{eqnarray}
The partonic level process is mediated by the SM $\gamma$ or $Z$ boson 
and the graviton. Interestingly, the interference between the SM and 
graviton diagrams identically vanish when the phase space integration 
is performed \cite{us1}.   To NLO the following subprocesses involving 
graviton contribute
\begin{eqnarray}
&&q + \bar q \rightarrow G  + g\,, \quad \quad q+\bar q \rightarrow G + 
\mbox{one~~loop}\,,
\nonumber \\[2ex]
&&q + g \rightarrow G  + q\,,
\qquad \bar q + g \rightarrow G  + \bar q\,,
\label{eq29}
\\[2ex]
&&g + g \rightarrow G + g \,, \quad \quad g+ g \rightarrow G+ 
\mbox{one~~loop}\,.
\nonumber 
\end{eqnarray}
The cross sections beyond LO involve the computation of one loop virtual 
gluon corrections and real gluon bremsstrahlung contributions to LO 
processes. We also include processes with a gluon in the initial state. 
Since we are dealing with energy momentum tensor coupled to gravity 
expressed in terms of renormalised fields and masses, there is no overall 
ultraviolet renormalisation required.  In other words, the operator 
renormalisation constant for the energy momentum operator is identical to 
unity to all orders in perturbation theory.  But we encounter soft and 
collinear divergences in our computation. These divergences are regulated 
using the dimensional regularisation.  We define $n=4+\varepsilon$ where 
$n$ is the space-time dimension.  With this procedure all divergences 
appear as $1/\varepsilon^\alpha$ where $\alpha=1,2$.  The soft divergences 
coming from virtual gluons and bremsstrahlung contributions cancel exactly 
according to the Bloch-Nordsieck theorem.  The remaining collinear 
divergences are removed by mass factorisation which is performed in 
the $\overline {MS}$ scheme.  The Drell-Yan coefficient function after mass 
factorisation, denoted by $\Delta_{ab}^i(z,Q^2,\mu_F^2)$, is computed by
\begin{eqnarray}
\bar \Delta_{ab}^i(z,Q^2,1/\varepsilon)=\sum_{c,d}
\Gamma_{ca}(z,\mu_F^2,1/\varepsilon) 
\otimes \Gamma_{db}(z,\mu_F^2,1/\varepsilon) \otimes 
\Delta_{cd}^i(z,Q^2,\mu_F^2)\,,
\label{eq30}
\end{eqnarray}
where $\bar \Delta_{ab}^i(z,Q^2,1/\varepsilon)$ is the bare partonic
coefficient function before mass factorisation is carried out.  The 
factorisation scale is given by $\mu_F$ and the kernel $\Gamma_{cd}(z)$ 
in the ${\overline {MS}}$ scheme is given by
\begin{eqnarray}
\Gamma_{cd}(z,\mu_F) &=&\delta_{c d} \delta(1-z)+ a_s {1 \over \varepsilon}
\left({\mu_F^2 \over \mu^2}\right)^
{\varepsilon/2} P_{cd}^{(0)}(z)\,,
\label{eq32}
\end{eqnarray}
where $a_s={\alpha_s(\mu_R^2) /4 \pi}$ and $P_{cd}^{(0)}(z)$ are the leading 
order Altarelli-Parisi splitting functions.  Expanding Eq.~(\ref{eq30}) up to 
order $a_s$, one can compute the coefficient function $\Delta_{ab}^i(z,Q^2,
\mu_F)$ from the bare $\overline \Delta_{ab}^i (z,Q^2,\mu_F,1/\varepsilon)$ 
and the known Altarelli-Parisi kernels $P_{ab}^{(0)}$. Finally we have to fold 
these finite $\Delta_{ab}^i (z,Q^2,\mu_F)$ with the appropriate partonic 
distribution functions to arrive at the $Q^2$ distribution for the DY pair. 
For completeness we present the results for the beyond SM effects coming due 
to the graviton exchange
\begin{eqnarray}
2 S~{d \sigma^{P_1P_2} \over dQ^2}(\tau,Q^2)&=&
\sum_q{\cal F}_{G} \int_0^1~ {d x_1 }~ \int_0^1 
~{dx_2}~ \int_0^1~ dz~ \delta(\tau-z x_1 x_2)
\nonumber\\[2ex]
&\times& \Bigg[ H_{q \bar q}(x_1,x_2,\mu_F^2) \Big(
\Delta_{q \bar q}^{(0),G}(z,Q^2,\mu_F^2)
 +a_s \Delta_{q \bar q}^{(1),G}(z,Q^2,\mu_F^2)\Big)
\nonumber\\[2ex]
&+&\Big( H_{q g}(x_1,x_2,\mu_F^2)+H_{g q}(x_1,x_2,\mu_F^2)\Big)
 a_s \Delta_{q g}^{(1),G}(z,Q^2,\mu_F^2) 
\nonumber\\[2ex]
&+& H_{g g}(x_1,x_2,\mu_F^2) \Big(
\Delta_{g g}^{(0),G}(z,Q^2,\mu_F^2)
 +a_s \Delta_{g g}^{(1),G}(z,Q^2,\mu_F^2)\Big) \Bigg]\,,
\label{eq36}
\end{eqnarray}
the factors ${\cal F}_{G}$ correspond to pure gravity part, $H_{ab}$ are 
the renormalised parton distributions and $\Delta_{ab}^{(0,1)}$ are given 
in \cite{us1}.

\section{Discussion}

The NLO-QCD corrections have been calculated in the ADD case for various 
distributions and for forward-backward asymmetry of the invariant lepton pair 
at both LHC and Tevatron \cite{us1}.  It was further extended to the RS 
model in \cite{us2}.  The un-integrated distribution with respect to $\cos 
\theta^*$ (the angle of a lepton with respect to the hadron in the {\it 
c.m} frame of the lepton pair) is particularly important as the interference
between the SM and gravity is not zero any more \cite{us3}.  
\begin{figure}[htb]
\centerline{
\epsfig{file=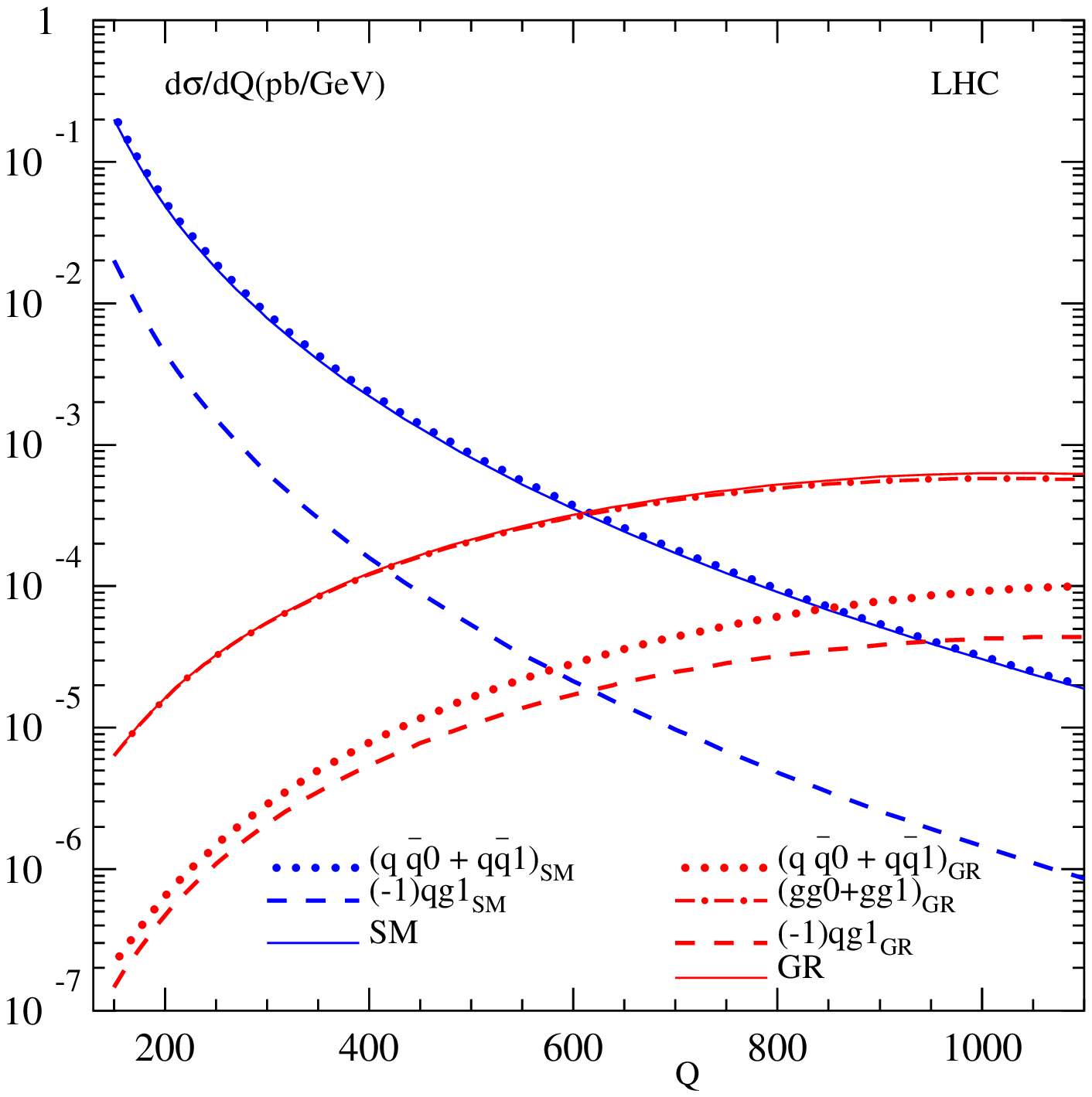,width=8cm,height=8cm,angle=0}
\epsfig{file=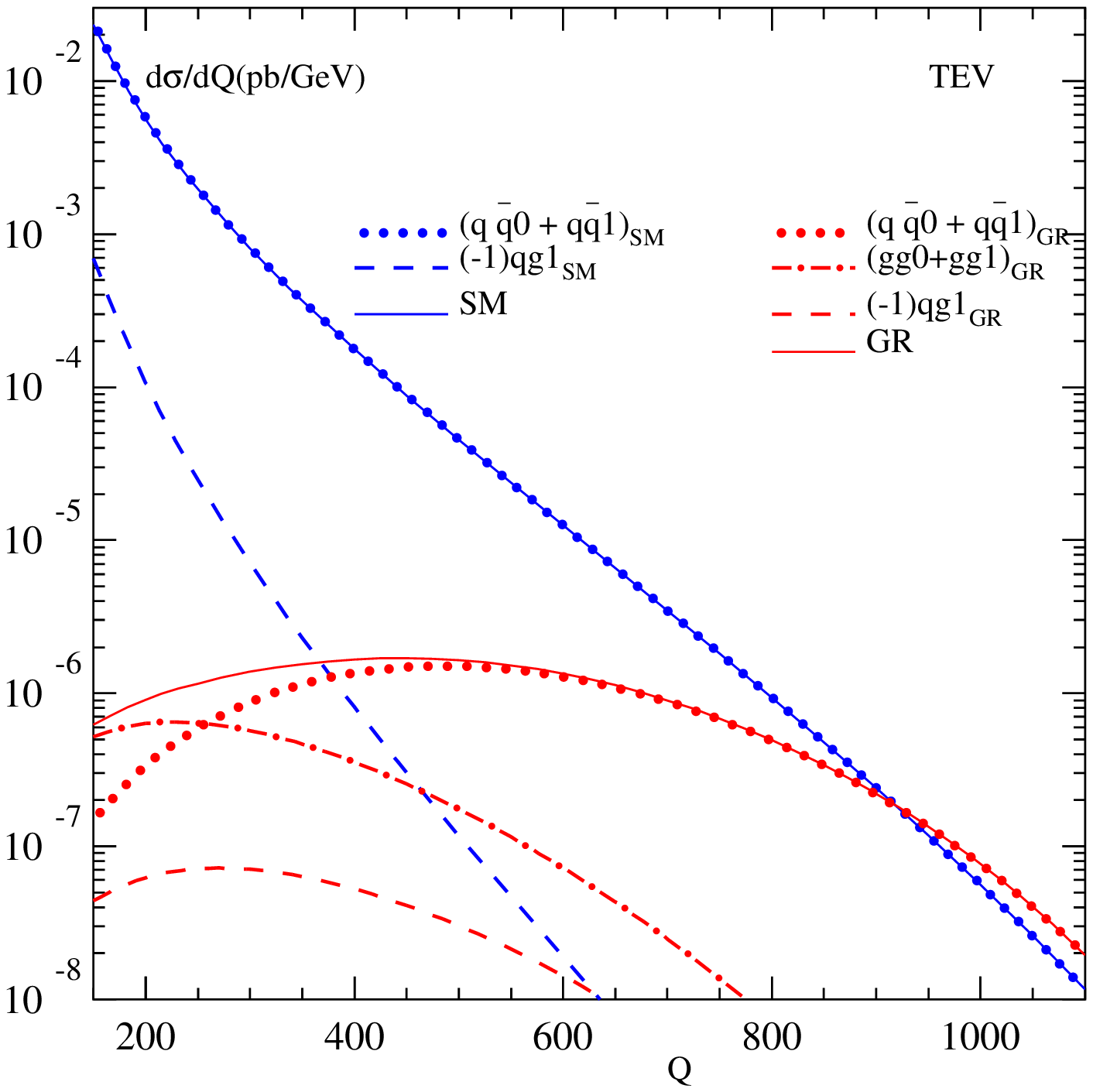,width=8cm,height=8cm,angle=0}}
\caption{The various contributing sub-process to the invariant mass 
distribution of the dilepton pair for ADD model at LHC and Tevatron.
}
\label{invQ}
\end{figure}
The inclusion of the higher order QCD corrections reduces the dependence on the factorisation 
scale significantly making the theoretical predictions for the relevant
observable more reliable for physics studies at high energies in hadron colliders.  
The dominant production mechanism for di-lepton production is through
gluons in the initial state in the case of LHC
and quarks in the case of Tevatron.  The parton distribution functions corresponding
to these initial states at LHC energies usually have large factorisation scale dependence.
We find that our results on NLO QCD corrections to the observable mentioned in this
paper do reduce the factorisation scale dependence coming from the parton distribution
functions as expected.  We also find that
the NLO QCD corrections quantified by the K factor are not small.    
The K-factor at large invariant mass is found to be as large as 1.6 at the LHC 
while at the Tevatron, it is similar to the SM value \cite {us1}.  
This large value at LHC is again due to the initial state gluons which have large 
flux at LHC energies.  The behavior
of the K-factor for RS model can be understood from the fact that only in
the resonance region the gravity part contributes, so at off resonance it is  
the SM K-factor while on resonance it is dominated by the gravity part.

\begin{figure}[htb]
\centerline{
\epsfig{file=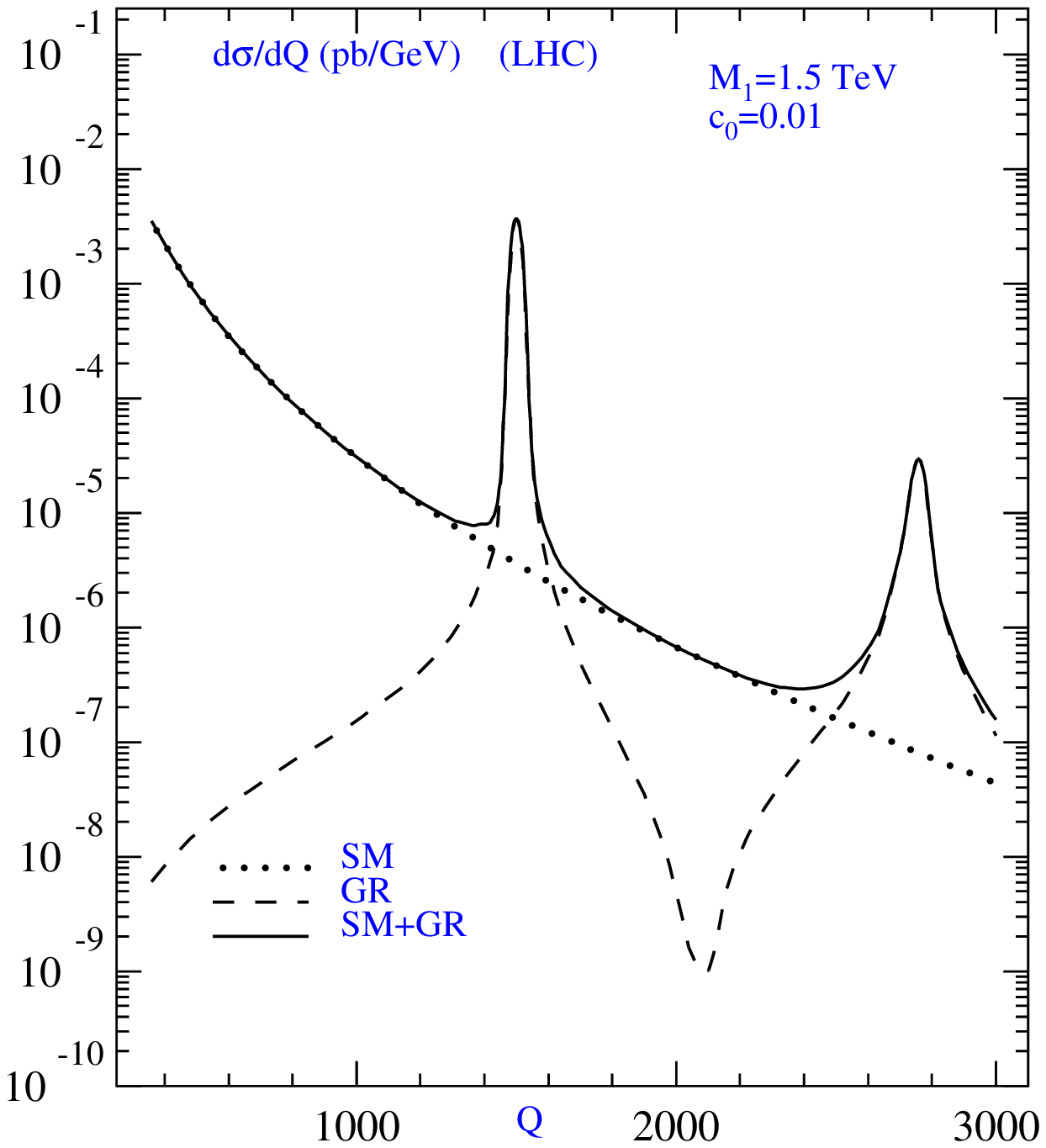,width=8cm,height=8cm,angle=0}
\epsfig{file=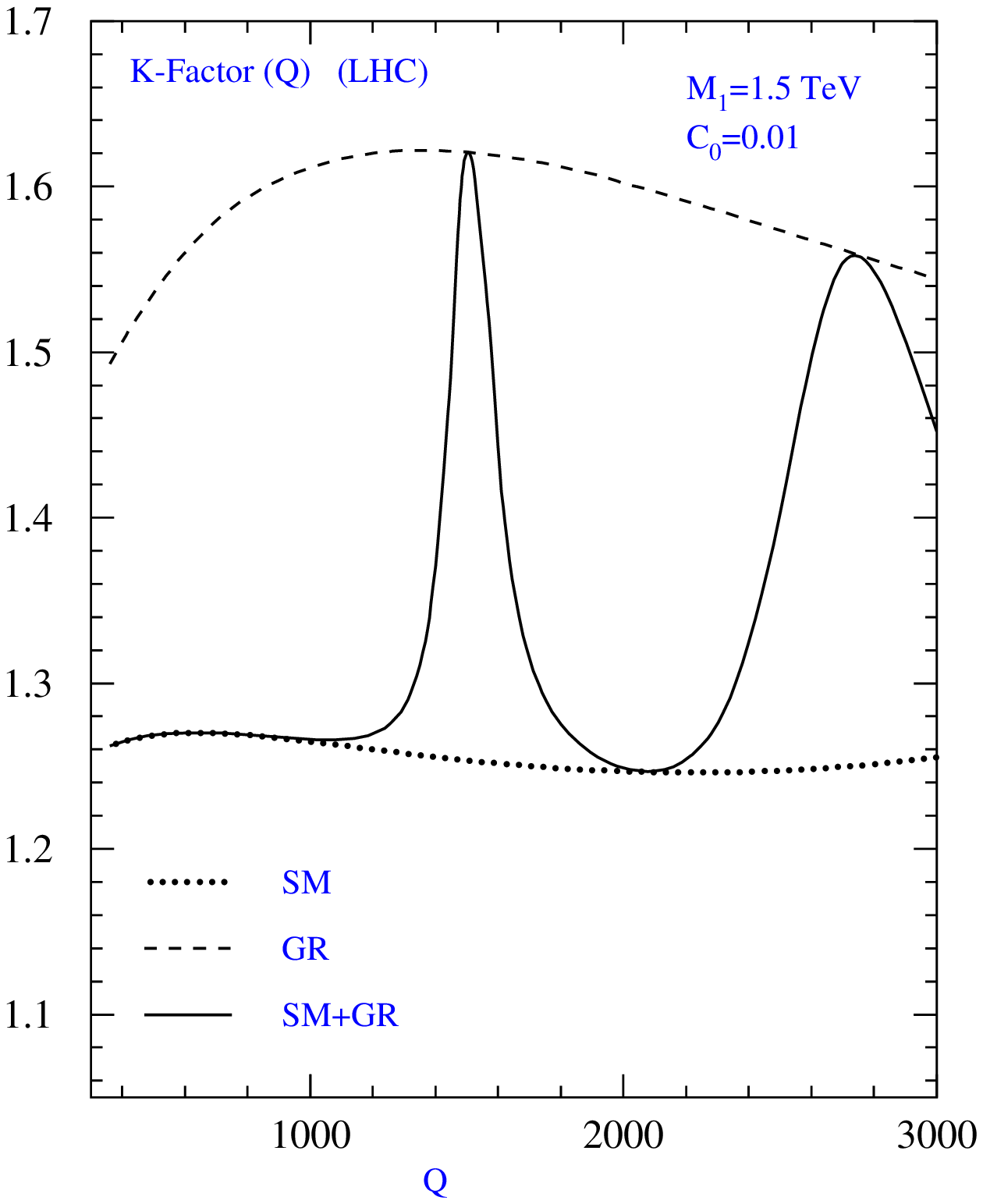,width=8cm,height=7.3cm,angle=0}}
\caption{Invariant mass distribution of the dilepton pair for RS model 
at LHC and the corresponding K-factor.
}
\label{rs}
\end{figure}
%


\end{document}